\documentclass[aps,nofootinbib,a4paper,floats,eqsecnum,12pt,tightenlines,preprintnumbers]{revtex4}

\usepackage[dvips]{graphicx}
\begin{document}
    
\newcommand{\beq}{\begin{equation}} 
\newcommand{\eeq}{\end{equation}} 
\newcommand{\beqa}{\begin{eqnarray}} 
\newcommand{\eeqa}{\end{eqnarray}} 
\newcommand{\da}{^\dagger} 
\newcommand{\wh}{\widehat}

\newcommand{\Om}{\Omega}
\newcommand{\om}{\omega}
\newcommand{\tr}{{\rm tr}}
\newcommand{\intf}{\int_{-\infty}^\infty}
\newcommand{\into}{\int_0^\infty}
\newcommand{\I}{{\mathcal I}}
\newcommand{\G}{{\mathcal G}}
\newcommand{\A}{{\mathcal A}}
\newcommand{\F}{{\mathcal F}}
\newcommand{\C}{{\mathcal C}}
\newcommand{\x}{\vec x}
\newcommand{\X}{\vec X}
\newcommand{\q}{\vec q}

\renewcommand{\=}{\!\!=\!\!}

\def\simleq{\; \raise0.3ex\hbox{$<$\kern-0.75em
      \raise-1.1ex\hbox{$\sim$}}\; }
\def\simgeq{\; \raise0.3ex\hbox{$>$\kern-0.75em
      \raise-1.1ex\hbox{$\sim$}}\; }

\def\la{{\langle}} 
\def\ra{{\rangle}} \def\vep{{\varepsilon}} \def\y{\'\i}
\def\half{{1\over 2}}
\def\an{|\Phi _N \rangle }
\def\bn{|\Phi ' _{N'}\rangle }
\def\na{ \langle \Phi _N|}
\def\nb{ \langle \Phi'_{N'} |}

\def\ov{\over}
\def\non{\nonumber }
\def\beq{\begin{equation} }
\def\eeq{\end{equation} }
\def\beqa{\begin{eqnarray}}
\def\eeqa{\end{eqnarray}}
\def\del{\partial }
\def\D{\Delta}
\def\a{\alpha }
\def\am{\alpha^\mu }
\def\Xm{X^\mu}
\def\Xn{X^\nu}
\def\d{\textrm{d}}
\def\b{\beta}
\def\t{\tau}
\def\e{\epsilon}
\def\g{\gamma}
\def\s{\sigma}
\def\med{\frac{1}{2}}

\def\npb#1#2#3{{Nucl. Phys.} {\bf B#1} (#2) #3}
\def\plb#1#2#3{{ Phys. Lett.} {\bf B#1} (#2) #3}
\def\prd#1#2#3{{Phys. Rev.} {\bf D#1} (#2) #3}
\def\prl#1#2#3{{Phys. Rev. Lett.} {\bf #1} (#2) #3}
\def\mpla#1#2#3{{Mod. Phys. Lett.} {\bf A#1} (#2) #3}
\def\ijmpa#1#2#3{{Int. J. Mod. Phys.} {\bf A#1} (#2) #3}
\def\jmp#1#2#3{{J. Math. Phys.} {\bf #1} (#2) #3}
\def\cmp#1#2#3{{ Commun. Math. Phys.} {\bf #1} (#2) #3}
\def\pr#1#2#3{{Phys. Rev. } {\bf #1} (#2) #3}
\def\umn#1#2#3{{ Usp. Matem. Nauk} {\bf #1} (#2) #3}
\def\bb#1{{\tt hep-th/#1}}
\def\grqc#1{{\tt gr-qc/#1}}
\def\heph#1{{\tt hep-ph/#1}} 
\def\mathph#1{{\tt math-ph/#1}}
\def\ptp#1#2#3{{\it Prog. Theor. Phys.} {\bf #1} (#2) #3}
\def\rmp#1#2#3{{\it Rev. Mod. Phys.} {\bf #1} (#2) #3}
\def\jetplet#1#2#3{{\it JETP Letters} {\bf #1} (#2) #3}
\def\jetp#1#2#3{{\it Sov. Phys. JEPT} {\bf #1} (#2) #3}
\def\pha#1#2#3{{\it Physica } {\bf A#1} (#2) #3 }
\def\app#1#2#3{{\it Astropart. Phys.} {\bf #1} (#2) #3 }
\def\cqg#1#2#3{{\it Class. Quantum Grav.} {\bf #1} (#2) #3 }
\def\apj#1#2#3{{\it Astrophys. J.} {\bf #1} (#2) #3 }
\def\ma#1#2#3{{\it Math. Ann.} {\bf #1} (#2) #3 }
\def\npps#1#2#3{{\it Nucl. Phys. Proc. Suppl.} {\bf #1} (#2) #3 }
\def\atmp#1#2#3{{\it Adv. Theor. Math. Phys.} {\bf #1} (#2) #3 }
\def\jhep#1#2#3{{ J. High Energy Phys.} {\bf #1} (#2) #3}
\def\cm#1{{\tt cond-mat/#1}}
\def\aph#1{{\tt astro-ph/#1}}
\def\nc#1#2#3{{\it Nuovo Cimento } {\bf #1} (#2) #3 }
\def\nat#1#2#3{{\it Nature} {\bf #1} (#2) #3 }
\def\ijtp#1#2#3{{\it Int. J. Theor. Phys.} {\bf #1} (#2) #3 }
\def\prep#1#2#3{{\it Phys. Rep.} {\bf #1} (#2) #3 }
\def\aj#1#2#3{{\it Astron. J.} {\bf #1} (#2) #3 }


\title{Portrait of the String as a Random Walk}


\author{Juan L. Ma\~nes}
\affiliation{Departamento de F\y sica  de la Materia Condensada\\
 Universidad del Pa\'{\i}s Vasco,
 Apdo. 644, E-48080 Bilbao, Spain  \\
 {\bf wmpmapaj@lg.ehu.es}}


\date{\today}

\begin{abstract}
In this paper  we analyze a Rutherford type experiment where light probes are 
inelastically 
scattered by an ensemble of excited closed strings, and use the corresponding  
cross section to extract density-density correlators 
between different pieces of the target string. We find a wide 
dynamical range where the space-time evolution of typical 
highly excited closed strings is 
accurately described as a convolution of  brownian motions. Moreover, 
we show that if we want to obtain the same cross section by coherently 
scattering probes off a classical background, then this background 
has to be time-dependent and singular. 
This provides an example where  singularities arise, not 
 as a result of strong gravitational self-interactions, but as a 
 byproduct of the decoherence implicit in  effectively describing  the string 
 degrees of freedom as a classical background.

\end{abstract}


\maketitle

\section{Introduction}

The idea that highly excited fundamental strings behave like random walks is not 
a new one. It was introduced as a physically sensible assumption 
in~\cite{sal}, where a model to describe a gas of strings at very 
high density was proposed. Namely, it was assumed that at fixed time 
a long string resembles the path of a brownian motion, so that the 
average distance between two points separated by a length $l$ of 
string is proportional to $\sqrt{l}$.
This idea is supported by   the dependence of the mean 
square radius of the string on its mass, $\la r^2\ra\propto M$, 
which is assumed to be proportional to the length. This 
was first obtained by oscillator methods by the authors of~\cite{mitp,mit}, 
who also    applied their techniques   to the study of cosmic string networks.

More recently, the random walk picture has been  useful  in different 
situations where  
highly excited strings are involved. It was used 
in~\cite{thor1,thor2} to derive the form of the interactions in a 
Boltzmann equation approach to the Hagedorn gas, and 
in~\cite{khuri} to model the string-black hole phase 
transition that  takes place at the correspondence 
point~\cite{cor}. In fact, the use of polymer physics techniques 
in~\cite{khuri}
yields a description of the string-black hole transition largely 
consistent with the one obtained in~\cite{hor} by a thermal scalar 
formalism and in~\cite{dam} by more conventional oscillator methods.
A discussion of the relation between the thermal scalar and the 
counting of 
random walks  has been recently given  in~\cite{pepe}, where the random walk 
is  used as an intuitive geometrical picture  in order to explain 
many features of the Hagedorn regime.
 Also closely related to the random walk picture is the approach where the 
string is actually assumed to be a composite of elementary 
`string-bits'~\cite{thorn, brane, halyo, rama}.

In spite of the undeniable usefulness of the idea, most of the results 
obtained from the random walk approach have been of a qualitative or 
at best semi-quantitative nature. One reason is  that the relation 
between excited strings and random walks has never been made very 
precise, and it is not hard to see why. The random walk idea is 
basically a statement about the size and shape of highly excited 
strings but, as mentioned above, the only 
\textit{geometric} properties  actually computed for the string and compared with the
  random walk predictions are  mean square radius~\cite{mitp, mit}. 
  And obviously, 
  a distribution is hardly characterized by its second 
moment alone.

Trying to refine the geometric description of  a  string state is tricky, 
since the 
coordinates $X^\mu$ are quantum fields on the worldsheet and, as 
such, undergo infinite zero-point fluctuations~\cite{had}. Indeed, even 
the mean square radius turns out to be infinite, and the above 
mentioned dependence on the mass of the string is obtained only after 
a subtraction is performed by hand. Attempts to obtain the 
higher moments $\la r^n\ra$ of the distribution directly by oscillator methods will
involve increasingly  complicated computations and an endless 
sequence of ambiguous, infinite 
subtractions.

An alternative approach is based 
on the observation~\cite{size} that we get infinite results because we 
sum aver the infinite modes of the string, whereas any real attempt 
to measure its shape  will be limited by the time 
resolution $\e$ of the experiment. Thus,  all modes with frequency 
$w>1/\e$ will be averaged out and effectively cut off. In particular 
one can use scattering experiments, where the time resolution is 
related to the energy by $\e\sim 1/E$. The cross sections can be used 
to extract form factors, and these are finite without the need for 
infinite subtractions and yield detailed information about the 
geometry of the string.
This approach has been used~\cite{ lor,comp} to study the energy 
dependence of the size of string ground  states, with surprising 
consequences regarding the behavior under Lorentz transformations and 
the black hole complementarity principle. It has also been used for 
open strings on the leading Regge trajectory~\cite{bo}, and very 
recently for 
typical closed strings~\cite{ff}.

In this paper we  analyze the results of~\cite{ff}, where the 
cross section for a Rutherford type experiment with light strings 
 scattered off highly excited 
closed strings was obtained. Rather than considering 
individual 
excited states, the equivalent of an `unpolarized' cross section was 
computed by taking an average over all the  states at a given 
mass level. This is exactly the kind of information that we need in 
order to settle the issue of the equivalence between strings and 
random walks, since at most one should expect an ensemble of highly 
excited strings to behave like some collection of random walks, and 
this only for some dynamical range.

This paper is organized as follows. In section~2 we present a brief 
review of the effective form factors computed in~\cite{ff} and find 
the first indication of random walk behavior in their elastic part, together 
with a puzzle. 
This is solved in section~3, where the inelastic cross section is used 
to obtain the time dependence of density-density correlators for the 
target string, yielding a picture of the space-time evolution of 
typical strings in terms of a convolution of random walks. In 
section~4 we show that, if one tries to obtain the same cross sections by coherently 
scattering probes off a classical background, then this background 
has to be both time-dependent and singular. Finally, our conclusions 
and outlook are presented in section~5.

\section{Elastic form factors and random walks}

\subsection{Review of string form factors}

 A 
natural way to measure the geometry of a string is by a scattering 
experiment. Although string amplitudes are rather complicate even for the simplest 
scattering processes and reflect exchanges of 
an 
infinite number of states, a simplification takes place in the Regge limit of 
fixed momentum transfer $t=-q^2$ and high  energy $E$. In this region 
the interaction is dominated by $t$-channel exchange of leading Regge 
trajectory states.
Consider for instance the Virasoro-Shapiro amplitude $\A(s,t)$ describing 
the scattering of two closed string tachyons. In the Regge limit  
this amplitude can be written as~\cite{gsw,pol}
\beq\label{regge}
\A\sim s^{2-q^2}{\Gamma(\frac{1}{2}q^2-1)\over\Gamma(2-{1\over 
2}q^2)}
\eeq
where $s\sim 4 E^2$ and we have set $\a'=2$. 
At low momentum transfer the amplitude is dominated by the graviton 
(multiplet) pole 
in the gamma function and can be interpreted as the product of the 
high energy 
graviton interaction for point particles by 
the form factors $\F(q^2)$ of the scattered strings~\cite{comp}:
\beq\label{cregge}
\A\sim {E^4\over q^2}(e^{- q^2 \log E})^2={E^4\over q^2}\F^2(q^2)\,
\eeq

The same strategy is followed   in~\cite{ff}, where closed string 
tachyons\footnote{Tachyons are used as probes for the sake of 
simplicity. As explained in~\cite{ff}, using other light string 
probes introduces polarization dependent factors in the cross 
section, but the target form factor is unchanged.}
are used to probe a statistical ensemble of excited closed strings of mass $M$.  
An  averaged interaction rate is defined by \emph{averaging} over all initial target 
states at mass 
level $N$ and momentum $p$ and \emph{summing} over all final target states at mass 
level $N'$ and momentum $p'$. Concretely, 
\beq\label{avrate}
{\cal R}¥(N,N',k,k')\equiv {1 \over{\cal G}_{c}¥(N)}\sum_{ \Phi_i|_{N}}\sum_{ \Phi_f|_{N'}} 
     |\A¥(\Phi_i,\Phi_f,k,k')|^2\,,
 \eeq
where the masses of the initial and final states are given by  $M^2= 
2(N\!-\! 1)$ and \hbox{$M'^2= 2(N'\!\!-\!\!1)$} respectively,
${\cal G}_{c}¥ (N)$ is the degeneracy of the $N^\mathrm{th}$ closed 
string  mass level, 
and $k$ and $k'$ are the momenta of the incoming and outgoing 
tachyons. In the Regge limit of high energy probes ($E>>1$ in string units) 
and fixed momentum transfer \hbox{$q^2=(k+k')^2$} the interaction 
rate factorizes
\beq\label{closedregge}
{\cal R}¥(N,N',k,k')\sim  
M^4\left|{\Gamma(\frac{q^2}{2}-1)\over\Gamma(2-{q^2\over 
2})}(EE')^{1-{q^2\over 2}}\right|^2|\F_{NN'}(q^2)|^2\,,
\eeq
where $\F_{NN'}(q^2)$ is the  \emph{effective form factor} for the 
target string. For heavy targets, this is given by
\beq\label{fm}
\F_{NN'}(q^2)=M^{-q^2}\oint _{C_v} {dv\over v} v^{N-N'} \la V_0 (-q,1) V_0 
(q,v)\ra_{w}\,,
\eeq
which is valid up to relative corrections of order $O(1/M)$.
In this expression,  $V_{0}$ is the oscillator part of the 
\emph{open} string tachyon vertex 
operator with zero modes removed, $C_v$ is a contour satisfying 
$|w|<|v|<1$, and the correlator 
 is evaluated  on a cylinder with 
 modular parameter $w=e^{-\b}$, with $\b$ given by
 \beq\label{beta}
\b\approx \pi\sqrt{D-2\over 6 N}={1\over 2 M T_{H}}\;\;\; , \;\;\; 
T_{H}=\frac{1}{2\pi}\sqrt{\frac{3}{D-2}} 
\eeq
where $T_{H}$ is the Hagedorn temperature in $D$ space-time dimensions. 

The interpretation of $\F_{NN'}(q^2)$ as an effective form factor can be 
motivated by noting that, for elastic scattering ($N=N'$) and 
 $q^2\sim 0$, 
eq.~(\ref{closedregge}) becomes
\beq\label{fregge}
{\cal R}¥(N,N',k,k')\sim \left|{(ME)^2\over 
q^2}\F(q^2)\F_{NN}(q^2)\right|^2\;\;,
\eeq 
where we recognize  the square 
of the Regge 
limit of the Virasoro-Shapiro amplitude describing tachyon-tachyon 
scattering~(\ref{cregge}), with one tachyon form factor $\F(q^2)$ replaced by 
 $\F_{NN}(q^2)$. Nevertheless, the effective form factors are useful also 
for inelastic scattering $(N\!\neq \! N')$ and light targets. The reason is 
that~(\ref{closedregge}) holds as long as we are in the Regge limit 
of high energy probes ($E\to\infty$) and fixed momentum transfer 
$q^2$, where strings are known to interact by exchanging the  
leading Regge trajectory as a whole.

\subsection{Spatial distribution   and random walks}

Since we 
expect the random walk picture to be valid only for distances  larger 
than 
the string scale, we will assume $q^2\ll 1$. In this region 
the following simple expression for the form factor is valid
\beq\label{form}
\F_{NN'}(q^2)\approx {1\over 2\pi}\int_{0}^{2\pi} 
d\xi\exp\left[i (N'\!\!-\!\!N)\xi-{q^2\over 
2\b}\,\xi(2\pi-\xi)\right]\ 
\eeq
with relative corrections  of order 
$O(q^2)$.

A spatial distribution was obtained in~\cite{ff} by Fourier transforming the 
\emph{elastic}
form factor 
\beq\label{rho}
\rho(\x)={1\over (2\pi)^{d}}\int d^{d} q\, e^{i\vec q\cdot \vec x} 
\F_{NN}(q^2)={1\over 2\pi}\left({\b\over 2\pi}\right)^{{d\over 
2}}\int_0^{2\pi} 
d\xi h(\xi)^{-{d\over 2}} 
\exp\left(-{\b \vec x\,^2\over 2 h(\xi)}\right)\ ,
\eeq
where $h(\xi)\equiv\xi(2\pi-\xi)$ and $d\equiv D-1$. 
Although $\rho(\x)$ can not be evaluated analytically, it is obvious 
from~(\ref{rho}) that
\beq\label{norm}
\int d^{d}x\rho(\x)=1\,.
\eeq

We will now show that the density $\rho(\vec x)$ admits  
a simple interpretation in terms of random 
walks. Consider a collection of discretized random paths  $\x(\xi)$, 
where $\xi$ is proportional to  the number of steps, and such that 
$\x(0)=0$. It is well known~\cite{par} that, in the continuum limit, the 
 probability (density) for the walk to visit a point $\x$ is given by
\beq
P(\x,\xi)=(4\pi a \xi)^{-d/2} \exp	\Big(\!-{\x\,^2\over 4a\xi}\Big)
\eeq
where $a$ is the diffusion coefficient if $\xi$ is interpreted as  
`time' for a brownian motion. Now, imagine a collection of 
\emph{closed} random paths subject to the constraint $\x(0)=\x(2\pi)=0$. Then, the 
conditional probability for the walk to visit a point $\x$  
 is given by
\beq\label{rw}
P_{c}¥(\x,\xi)=A\,\exp	\Big(\!-{\x\,^2\over 4a\xi}\Big)\,\exp	
\Big(\!-{\x\,^2\over 4a(2\pi-\xi)}\Big)=A\,\exp\Big(\!-{\pi\x\,^2\over 
2a\xi(2\pi-\xi)}\Big)
\eeq
where the constant $A=[2a\xi(2\pi-\xi)]^{-d/2}$ normalizes the 
probability  $P_{c}¥(\x,\xi)$ to $1$. Then, the 
probability that the closed random walk visits a point $\x$ for 
\emph{any} value of $\xi$ is 
\beq\label{rw1}
P_{c}¥(\x)={1\over 2\pi}\int_{0}^{2\pi} d\xi P_{c}¥(\x,\xi)\;.
\eeq
But this is exactly  $\rho(\x)$ in~(\ref{rho}), with 
$a=\pi/\b$. In other words, $\rho(\x)$ is created by a collection of 
closed random walks that begin and end at the origin! Note that  we have 
recovered not just the  second moment $\la \x\,^2\ra$, but the complete 
(closed) random walk distribution from the cross section. And since one can 
show, by using 
the exact formula for $\F_{NN'}(q^2)$ given in~\cite{ff}, that the corrections to~(\ref{form})  
are rather  small 
for $q^2\simleq 1$, we may conclude that the string `looks' like a random 
walk even at resolutions close to the string scale.  
   
There is, however, something rather puzzling in this picture. 
According to~(\ref{rw1}), all the random paths representing the ensemble 
of excited closed strings intersect at  the origin. This is strange, 
since the origin is the 
center of mass, and there is no reason why  the strings should  go 
through it.
Closely related to this is the fact, 
pointed out in~\cite{ff}, that for $1\simleq r\ll \sqrt{M}$, 
$\rho(r)\propto  r^{2-d}$, i.e. the string distribution is singular at 
the origin. This would be quite natural for self-gravitating strings, 
but   is rather surprising for the ensemble of \emph{free} 
strings considered here. We will tackle these problems in the next section.

\section{Inclusive scattering and structure functions}
\subsection{Density-density correlators}
In order to better understand the physical meaning of the density 
$\rho(\x\,)$, 
it is convenient to consider the scattering cross section associated 
to the averaged rate~(\ref{avrate}). This is obtained by adding the 
appropriate phase factors and closed string coupling constant 
$g_{c}$. In the limit $M\gg E\gg 1$, where target recoil can be 
neglected and the 
the center of mass 
frame
coincides with the target rest frame, the cross section is given by
\beq\label{crossM}
{d\sigma\over d\Omega}={g_{c}^4\over 
16 E M^2}{ k'^{D-3}\over (2 \pi)^{D-2}} {\cal R}¥(N,N',k,k')\,. 
\eeq
For elastic scattering ($N\!=\!N',\;k\!=\!k'\!=\!E$) and $q^2\ll 1$, we can use the 
approximate expression~(\ref{fregge}) for the rate, and the cross 
section becomes
\beq\label{born}
{d\sigma\over d\Omega}={g_{c}^4 \over 
16  (2 \pi)^{D-2} }\left|E V(\q\,) \F(\q\,^2)\right|^2  k^{D-2}
\eeq
where we have defined a potential
\beq
V(\q\,)=M{\F_{NN}(\q\,^2)\over \q\,^2}
\eeq
or, in terms of the Fourier transform~(\ref{rho}),  $\Delta V(\x)=-M \rho(\x)$. 
Except for the presence of the tachyon form factor $\F(\q\,^2)$ that 
reminds us that the probes are not point-like,
(\ref{born}) is just the formula for Born scattering\footnote{Note that 
the phase space factor $k^{D-2}$ in~(\ref{born}) is 
appropriate for massless probes. In  the more familiar case of nonrelativistic  scattering
with probes of mass $m$, the phase factor would be replaced by $mk^{D-3}$.}  
 by a newtonian potential $V(\x\,)$.
In other words, the experimenter could interpret the elastic 
 data as due to scattering by a target with mass distribution $M\rho(\x\,)$.

For truly elastic scattering, where not only the energy but also the 
state of the target string is unchanged $\Phi_{f}=\Phi_{i}$, 
$\rho(\x\,)$ would represent  the density of the target string. In 
terms of Fourier transforms,
\beq\label{naive}
 \F_{NN}(\q\,^2)=\la \Phi_{i}¥ |\rho_{st}( \q\,)|\Phi_{i}\ra\, ,
\eeq
where $\rho_{st}$ is a suitably defined string density operator. 
 However, we are considering an inclusive process where one averages 
 over  all initial states with fixed mass $M$ and momentum $p$, and sums over 
  final states of mass $M'$ and momentum $p'$. Thus, even for 
  `elastic' ($M\!=\!M'$) scattering, the na\"\i ve relation~(\ref{naive}) does not hold and 
  $\rho(\x\,)$, as given by~(\ref{rho}), can \emph{not} be understood as 
  the actual  density of 
 the target string. Instead, the 
  formula for the averaged inclusive rate~(\ref{avrate}) implies
\beqa\label{double}
|\F_{NN'}(q^2)|^2&=&{1 \over{\cal G}_{c}¥(N)}\sum_{ \Phi_i|_{N}}\sum_{ \Phi_f|_{N'}} \big|
 \langle \Phi_f |  \rho_{st}(\q\,)|\Phi_i \rangle \big|^2\non \\
 &=&{1 \over{\cal G}_{c}¥(N)} \sum_{ \Phi_i|_{N}}\sum_{ \Phi_f|_{N'}} \langle 
 \Phi_i |   \rho_{st}(-\q\,)|\Phi_f \rangle 
 \langle \Phi_f |  \rho_{st}(\q\,)|\Phi_i  \rangle\;. 
\eeqa
Note that we have dropped the elastic scattering constraint 
$N\!=\!N'$ as  we are  considering a general situation.
In order to proceed, it is convenient to introduce the following 
function
\beq\label{cn}
\C_{N}(\vec q\,,\om)\equiv\sum_{N'}¥|\F_{NN'}(q^2)|^2\delta (M'-M-\om)
\eeq
where $\om\equiv q^0$ is the energy transfer. Then, writing   the delta function 
as a  Fourier transform  
\beq
\delta (M'-M-\om)={1\over 2\pi}\int_{-\infty}^\infty d(t-t') 
e^{i(M'-M-\om)(t-t')}
\eeq
and using 
\beq
\la \Phi_{f}¥ |\rho_{st}(\q\,)|\Phi_{i}\ra e^{i(M'-M)t}=\la \Phi_{f}¥ 
|\rho_{st}(\q\,,t)|\Phi_{i}\ra\;,
\eeq
the double sum in~(\ref{double}) can be converted into a time 
dependent density-density correlator
\beq
\C_{N}(\q\,,\om)={1\over 2\pi}\int_{-\infty}^\infty d(t-t') 
e^{-i\om(t-t')}\la \rho_{st}(-\q\,,t')\rho_{st}(\q\,,t)\ra_{N}
\eeq
where the  ensemble average $\la A\ra_{N}¥$ of an operator  is defined by 
\beq\label{ens}
\la A \ra_{N}¥\equiv {1 \over{\cal G}_{c}¥(N)}\sum_{ \Phi_i|_{N}}\la \Phi_{i}| A | 
\Phi_{i}\ra \, .
\eeq

$\C_{N}(\vec q\,,\om)$ is thus the dynamical \emph{structure 
function}, i.e., 
the Fourier transform of the time-dependent density-density correlator 
for the ensemble of target strings. Inverting the Fourier transforms 
gives
\beq\label{foucn}
\int d^d x\la\rho_{st}(\x\,+\Delta\x\,,t+\Delta t) \rho_{st}(\x\,,t)\ra_{N}={1\over (2\pi)^d}\int 
d^d q\, 
d \om\,
e^{i(\q\,\cdot \Delta\x-\om\Delta t)}\, \C_{N}(\q\,,\om)
\eeq 
where $d=D-1$ and $(\Delta\x,\Delta t)$ is an arbitrary, not necessarily small, 
space-time displacement. Finally, using~(\ref{cn}) yields the correct 
relation between the string density $\rho_{st}$ and the form factor 
$\F_{NN'}$
\beq\label{cor}
\int d^d x\la \rho_{st}(\x\,+\Delta\x\,,\Delta t)\rho_{st}(\x\,,0)\ra_{N}=\sum_{N'}
{1\over (2\pi)^d}\int d^d q \,
e^{i(\q\,\cdot \Delta\x-\om\Delta t)}\, |\F_{NN'}(q^2)|^2
\eeq
where $\om=M'-M$ and we have set $t=0$ since, according to~(\ref{foucn}), the correlator 
depends only on the time difference $\Delta t$. Thus the (partially integrated) 
density-density correlator 
is the Fourier transform of the square of the inelastic form factor.
Actually, it is  a series rather than a Fourier transform in $\om$, 
implying  periodicity in time with the  period $T$ fixed by the mass shell 
condition 
\beq
\om=M'\!-\!M\simeq  (N'\!\!-\!\!N)/M\;\;\Longrightarrow\;\; T=2\pi M\;.
\eeq

\subsection{Probabilities from correlators}

The Fourier transform of the form factor is easily obtained by noting
that the spatial part is a simple Gaussian 
and~(\ref{form}) is already given as a  Fourier coefficient 
with  \hbox{$i (N'-N)\xi=i \om t$}. The  result is 
\beq\label{fouform}
\sum_{N'}{1\over (2\pi)^d}\int d^d q \,
e^{i(\q\,\cdot \x-\om t)}\, \F_{NN'}(q^2)\simeq (\pi 
f(t))^{-{d\over2}} \exp\Bigl(-{\x\,^2\over f(t)}\Bigr)\equiv P_{c}(\x,t) 
\eeq
where $f(t)$ is a periodic function that, for $0<t<T\!=2\pi M$, is given by
\beq\label{ft}
f(t)={4 T_{H}\over M}t\,(T-t)\;.
\eeq

This shows that  the Fourier 
transform of the inelastic form factor is just the conditional 
probability~(\ref{rw}) with $t=M\xi$.
In evaluating~(\ref{fouform}), the approximation 
\hbox{$q^2=\q\,^2-\om^2\simeq 
\q\,^2$} has been made. This is justified by noting that, at low 
momentum transfer,  (\ref{form}) is 
non-negligible only for $\om\simleq \q\,^2$. 
By~(\ref{cor}), the density-density correlator can be written as a 
space-time convolution of  conditional probabilities
\beq\label{conv}
\int d^d x\la \rho_{st}(\x\,+\Delta\x\,,\Delta t)\rho_{st}(\x\,,0)\ra_{N}=
{1\over T}
\int_{0}^{T}¥  d t'\int d^d x' \,P_{c}(\x',t') \,P_{c}(\Delta\x-\x',\Delta 
t-t')\;.
\eeq

The physical meaning of this equation, which is the main result of this section,
can be clarified by noting that the density-density correlator
\beqa
\la \rho_{st}(\x\,+\Delta\x\,,\Delta t)\rho_{st}(\x\,,0)\ra_{N}\non
\eeqa
is proportional to the conditional probability to detect a string 
(with center of mass at the origin) at 
point $\x\,+\Delta\x$ for  $t=\Delta t$, assuming  the string  has been previously 
detected at point $\x$ for $t=0$. This probability is \emph{not} 
translational invariant and depends not only 
on the distance $\Delta\x$, but also on $\x$. This is so because the ensemble includes 
all states at mass level $N$ with fixed total momentum $p$, i.e., we  
average over oscillator states, but not over center of mass wave 
functions. In position space, our target is a highly excited string 
with center of mass at the origin, and the correlator has to go 
to zero as its arguments move away from it. 

Integrating over 
$\x$ is obviously equivalent to dropping the fixed center of mass 
constraint. In other words, the integrated correlator in the l.h.s. of
~(\ref{conv}) is simply proportional to the probability $P_{st}(\Delta\x\,,\Delta t)$ 
that \emph{any} string of mass 
$M$ that has been previously detected at a point is detected again after a time $\Delta t$ at 
a distance $\Delta\x$ , irrespective of its 
center of mass position. Actually, the constant 
of proportionality is~$1$ and~(\ref{conv}) can be rewritten as
\beq\label{proconv}
P_{st}(\Delta\x\,,\Delta t)={1\over T}
\int_{0}^{T}¥  d t'\int d^d x' \,P_{c}(\x',t') \,P_{c}(\Delta\x-\x',\Delta t-t')
\eeq
since, as one can easily verify, $P_{st}$ is correctly normalized
\beq 
\int d^d x P_{st}(\x\,, t)=1\;.
\eeq 
The periodicity in time mentioned above means that a configuration of  
the ensemble of target strings repeats itself after a time $2\pi M$:
\beq
P_{st}(\Delta\x\,,\Delta t+2\pi M)=P_{st}(\Delta\x\,,\Delta 
t)\;.
\eeq

For $\Delta t=0$, doing the $x'$ integral in~(\ref{proconv}) gives the 
probability for simultaneous detection of the string at two points 
separated by $\Delta \x$. The result is 
\beq\label{intfree}
P_{st}(\Delta\x\,,0)={1\over T}\int_{0}^{T}¥  d t' 
P_{c}(\Delta\x/\sqrt{2},t')={1\over 4\pi M}\int_{0}^{4\pi M}¥  d\s 
P_{c}(\Delta\x/ \sqrt{2},\s/ 2)
\eeq
where we have made the change of variables $\s=2t'$. 
Note that 
\beq\label{free}
P_{c}(\Delta\x/ \sqrt{2},\s/ 2)\propto \exp\bigg(-{M\over 
2T_H}{\Delta\x\,^2\over \s(4\pi M-\s)}\bigg)
\eeq
has the same form as~(\ref{rw}) and represents the probability that 
any closed path of length $4\pi M$ and unconstrained  center of mass  
goes simultaneously through  two 
points separated by an interval $\Delta\x$. Thus, at any given time 
the ensemble of strings looks like a 
collection of closed random walks parametrized by $0\le \s \le 4\pi 
M$.

Although~(\ref{intfree}) is formally similar to~(\ref{rho}) 
or~(\ref{rw1}), the  interpretation is quite different. While in~(\ref{rho}) 
or~(\ref{rw1}) $\x$ is the distance to a \textit{fixed} point, the 
center of mass, $\Delta\x$ in~(\ref{intfree})
 is the distance  to \textit{any} point where the string is 
simultaneously detected. Thus, recognizing that the string density 
$\rho_{st}$ is related to $\F_{NN'}$ by~(\ref{cor}), rather than 
by~(\ref{rho}), avoids the problematic conclusion that all the strings go through 
a common point and solves the puzzle mentioned at the end of last 
section.

Moreover, (\ref{proconv}) contains more than purely static 
information. Indeed, (\ref{proconv}) expresses the 
time evolution of the ensemble of highly excited strings as a 
convolution of (closed)  brownian motions. Concretely, the 
probability that any  string goes through two different points separated by a 
space-time interval $(\Delta\x,\Delta t$)  involves two closed random 
paths: one going from the initial point to any space-time point within 
a time period $T=2\pi M$, and another going from this intermediate point to 
the final point. In what follows, we will show that this somewhat 
surprising structure is quite natural from a semiclassical point of 
view.

\subsection{Semiclassical interpretation}
Any classical closed string solution~\cite{gsw,pol} can be written as the sum of two 
arbitrary functions subject  to the constraints 
$(\partial_{+}¥X_{\!+})^{\,2}=(\partial_{-}¥X_{\!-})^{\,2}=0$,
\beq
X^\mu(\t,\s)=X_{+}^\mu(\s_{+})+X_{-}^\mu(\s_{-})\;\;,\;\; \s_{\pm}=\med(\t\pm\s)
\eeq
where  $X^\mu(\t,\s+T_{\s}¥)=X^\mu(\t,\s)$. 
It is easier to  work  in the temporal gauge defined by
\beq\label{gauge}
X^0=\t\;\;,\;\; X_{\pm}^0=\s_{\pm}¥\;\;.
\eeq

The period $T_{\s}$ is determined by noting that the total energy-momentum 
of the string is given by the Noether current
\beq\label{noe}
P^\mu={1\over 4\pi}\int_{0}^{T_{\s}}d\s\,\partial_{\t}X^\mu\;.
\eeq
This formula is valid for a string tension 
$(2\pi\a')^{-1}=(4\pi)^{-1}$. In 
the center of mass frame, $P^{\,0}\!=\!M$ and~(\ref{noe}) implies 
$T_{\s}=2T=4\pi M$,
and $\vec X_{\pm}(\s_\pm+2\pi M)=\vec X_{\pm}(\s_\pm)$. The 
constraints become
\beq\label{cons}
(\partial_{+}¥\vec X_{+})^{\,2}=(\partial_{-}¥\vec X_{-})^{\,2}=1
\eeq
and we see that each classical solution is represented  by  two arbitrary 
$d$-dimensional closed curves of length $2\pi M$. On each curve, 
$\s_{\pm}$ plays the role of length parameter.

Now, the classical probability for a string to go through two points 
separated by the interval $(\Delta \x,\Delta t)$ will be proportional 
to the `number' of pairs of $d$-dimensional closed curves of length $2\pi M$ 
that go through the first point, that we may choose as the origin, 
\beq
\vec X_{\pm}(0)=\vec X_{\pm}(2\pi M)=0\;\;,
\eeq
and such that
\beq
\vec X_{+}(\s_{+})+\vec 
X_{-}(\s_{-})=\Delta\x\;\;,\;\;\s_{\pm}=\med(\Delta t\pm \s)
\eeq
for any value of the parameter $\s\!\in\![0,2T]$. If one assumes that the number of 
curves satisfying $\vec X_{\pm}(\s_{\pm})=\x$ is given by 
$P_{c}(\x,\s_{\pm})$, then the classical probability is necessarily 
given by
\beq
P_{st}(\Delta\x\,,\Delta t)={1\over 2T}
\int_{0}^{2T}¥  d \s\int d^d X_{+} \,
P_{c}\Big(\vec X_{+},\med(\Delta t+\s)\Big)P_{c}	\Big(\Delta\x-\vec 
X_{+},\med(\Delta t-\s)\Big)\;\;.
\eeq
With the substitutions $\s=2t'-\Delta t$ and $\vec X_{+}=\x\,'$, this 
is exactly~(\ref{proconv}).

This argument is heuristic because the `number of curves' is always 
infinite, unless a cut-off is introduced. In fact, this cut-off is 
forced upon us  by the breakdown of the classical description. A string 
can  be viewed as a collection of oscillators $\a^\mu_{n}$, that may be described 
classically only for large values of occupation numbers  $N_{n}^\mu$. The 
mass level is given by
\beq
\half M^2+1=N=\sum_{n=1}^\infty\sum_{\mu=1}^{D}\,n\,N_{n}^\mu
\eeq
Then, for large $N$, 
\beq
\la N_{n}^\mu \ra\sim {\sqrt{N}\over n}\sim {M\over n}
\eeq
and the oscillators admit a classical description only for $n$ less 
that the mass of the string in string units. Since we are considering curves of 
length $2\pi M$, the wavelength corresponding to the oscillator 
$\a^\mu_{n}$ is $\lambda\sim M/n$, and the breakdown takes place for 
$\lambda\sim 1$ , i.e., at the string scale. Thus, each curve should 
be discretized as a collection of $\sim M$ pieces or string bits. Note that in this 
argument the string bits appear, not as fundamental objects, but as an 
effective way of describing  a highly excited string semiclassically.

Now, counting curves with a cut-off $\e$ is equivalent to counting 
random walks, and in the continuum limit the probability that an 
ordinary random walk goes from the origin to $\x$ is~\cite{par} 
\beq
P(\x,\s)\sim \exp	\Big(\!-{b\x\,^2\over \e\s}\Big)
\eeq
as long as $\s$ is a length parameter. The constant $b$ depends both 
on the dimension $d$ and
the details of the discretization, but is of 
order one. For a closed random walk of length $T$
\beq
P_{c}¥(\x,\s)\sim\exp	\Big(\!-{b\x\,^2\over \e\s}\Big)\,\exp	
\Big(\!-{b\x\,^2\over \e(T-\s)}\Big)\sim\,\exp\Big(\!-{bT\x\,^2\over 
\e\s(T-\s)}\Big)
\eeq
This is~(\ref{fouform}) for $\e\sim 1$,  up to a constant of order one 
in the exponent.

\section{Decoherence and singular backgrounds}

In this section we will try to understand the differences between the 
string density $\rho_{st}¥(\x,t)$, related to the form factor 
by~(\ref{cor}), and the density $\rho(\x)$ given by~(\ref{rho}).
As argued at the beginning of section~3, the fact that the effective 
form factor is obtained from an averaged inclusive cross 
section implies that $\rho(\x)$ can \emph{not} represent the actual 
string density of the target. They are in fact very different 
functions. As mentioned above, $\rho(\x)$ is singular at the origin, 
with a divergence $\rho(r)\sim r^{2-d}$. On the other hand, according 
to the analysis in section~3, an instantaneous picture of the target would show an 
ensemble of closed random walks, all of them with center of mass at the origin. It 
is intuitively obvious that the associated density should be perfectly 
regular. 

This can be shown explicitly. Although~(\ref{intfree}) is missing the fixed 
center of mass constraint due to the $\x$-integral in~(\ref{cor}), 
this condition is easily restored with the use of a path integral 
formalism. The details of this computation are presented in the 
appendix, where it is shown that
\beq\label{rhost}
\rho_{st}¥\,(\x,t)=B\exp\bigg(\!-{3\x\,^2\over8\pi^2MT_H}\bigg)
\eeq
with $B=(8\pi^3MT_H/3)^{-d/2}$. Thus $\rho_{st}$ is independent of 
$t$, as one would expect of an ensemble of stationary states, and 
regular. The mean square radius computed  from it 
\beq
\la\, \x\,^2\ra={4\pi^2 d\over 3} T_{H}M
\eeq
agrees with the result obtained by oscillator methods~\cite{mitp,mit}.

The analysis at the beginning of  
section~3 shows that the elastic cross section could also be interpreted as 
due to potential scattering with $\rho(\x)$ acting as a source, i.e., 
$\Delta V(\x)=-M\rho(\x)$. However, this interpretation is  
incomplete because of the 
the occurrence of inelastic processes that can not be 
accounted for by a time-independent potential. We are thus led to 
consider the inelastic generalization of the cross section~(\ref{born}). Taking the 
$q^2\ll 1$ limit of~(\ref{closedregge}) and using~(\ref{crossM}) yields
\beq\label{inborn}
{d\sigma\over d\Omega}={g_{c}^4 \over 
16  (2 \pi)^{D-2} }\left|V(\q\,,\om) \F_{EE'}¥(\q\,^2)\right|^2  EE'\,k'^{D-2}
\eeq
where we have defined
\beq\label{tpot}
V(\q\,,\om)=M{\F_{NN'}(\q\,^2)\over \q\,^2}
\eeq
with $\om=E-E'=M'-M\simeq\sqrt{2N'}-\sqrt{2N}$. $\F_{EE'}$ can 
be interpreted as   an 
inelastic tachyon form factor and  is given by
\beq
\F_{EE'}¥(q^2)=e^{-{q^2\over 2}\log EE'}\;.
\eeq

Thus, the main differences between~(\ref{inborn}) and the elastic 
cross-section~(\ref{born}) are the replacements
\beq
|E\F(\q\,^2)|^2\longrightarrow EE'| \F_{EE'}¥(\q\,^2)|^2\;\;,\;\;  
V(\q\,)\longrightarrow V(\q,\om)\;.
\eeq
The first one  is quite trivial and reflects the fact that the 
incoming and outgoing probes have 
different energies. The second replacement means that, in order to have 
inelastic processes in potential scattering, the potential has to be 
time-dependent. The  Fourier transform of~(\ref{tpot}), $\Delta 
V(\x\,,t)=-M\rho_{eff}(\x\,,t)$, implies that the potential is created 
by the time-dependent effective density 
\beq\label{fourho}
\rho_{eff}¥ (\x\,,t)=\sum_{N'}{1\over (2\pi)^d}\int d^d q \,
e^{i(\q\,\cdot \x-\om t)}\, \F_{NN'}(\q\,^2)= (\pi 
f(t))^{-{d\over2}} \exp\Bigl(-{\x\,^2\over f(t)}\Bigr) 
\eeq
where $f(t)$ is the periodic function given by~(\ref{ft}). Then note  that, 
according to~(\ref{rho}), $\rho(\x)$ is simply the time average of 
$\rho_{eff}(\x,t)$
\beq\label{avrho}
 \rho\,(\x)={1\over T}\int_{0}^{T} dt\,\rho_{eff}¥ (\x\,,t)\;.
\eeq

The qualitative features of $\rho_{eff}¥ (\x,t)$ and the associated 
potential $V(\x,t)$ are rather 
suggestive. The  mean square radius of the distribution is 
\beq
\la\, \x\,^2\ra(t)={d\over 2} f(t)={2\,T_{H}d \over M}t\,(2\pi M-t)
\eeq
and goes to zero periodically for $t=n(2\pi M),\,n\!\in\! Z$, where  
the density collapses 
to a delta function.
Note that it takes a time $2\pi M$ for the radius to grow from zero to 
the maximum value \hbox{$r_{max}=¥\pi\sqrt{2MT_H d}$} and then collapse back to 
zero. Thus, for large $M$, the potential is 
slowly varying. This explains the absence of retarded effects in the 
relation \hbox{$V(\x\,,t)=-M\rho_{eff}(\x\,,t)$} and justifies  the use of an 
instantaneous newtonian potential\footnote{Technically, this is related to the 
approximation \hbox{$q^2=\q\,^2-\om^2\simeq \q\,^2$} used 
in~(\ref{inborn}) and~(\ref{tpot}). See also comment 
after~(\ref{ft}).}.

For $r>r_{max}$ the density goes to zero exponentially,  the potential 
approaches that of a point mass and tends to become  static.
Thus we have an outer, approximately static  region, and an 
inner dynamical region  where the potential is time-dependent  and corresponds to a 
spherically symmetric mass distribution that periodically collapses 
and bounces back. Note that this is vaguely reminiscent of the 
maximally extended Schwarzchild solution~\cite{grav}: Above the horizon the 
geometry is static but, inside the gravitational radius, $r$ is a 
time-like coordinate and the geometry becomes time-dependent, evolving from a 
`big bang' to a `big `crunch' in a time  proportional ---in four 
dimensions--- to the mass of the black hole.

Thus, from the same scattering cross sections we can get two entirely different
portraits of the scatterer. If we are aware of the decoherence 
implied by  averaging and 
summing over  enormous numbers of string states in~(\ref{avrate}), we will 
use~(\ref{cor}) to extract information on density-density 
correlators, ending up with a description of the target in 
terms of random walks. We will find that the `actual' density of the 
string $\rho_{st}$,  given by~(\ref{rhost}), is time-independent and  regular.
On the other hand, if we assume that the scattering is caused by a 
\emph{classical} background, we will conclude that the 
corresponding `apparent' or effective density $\rho_{eff}$ is both 
 time-dependent and singular.
 We may say that, at least in  the example considered here,  
 singularities arise not 
 as a result of strong gravitational self-interactions ---after all, 
 the target is a highly excited \emph{free} string--- but as a 
 byproduct of the decoherence implicit in  effectively describing  the string 
 degrees of freedom as a classical background.
 
 We would like to close this section by pointing out that $\rho_{eff}$,  given 
by~(\ref{fourho}), is not  unique. The reason is that the interaction 
rate~(\ref{closedregge})
determines the effective form factor only up to an arbitrary phase.  
Imposing time-reversal symmetry leads uniquely to the real  
form factor~(\ref{form}) and to the time-symmetric effective density~(\ref{fourho}). 
Dropping this constraint amounts to a redefinition
\beq\label{red}
\F_{NN'}(q^2)¥\to\F_{NN'}(q^2)\,e^{\,i\varphi(\q,\om)}\;,
\eeq
where reality of  $\rho_{eff}(\x,t)$ requires 
$\varphi(\q,-\om)\!=\!-\varphi(\q,\om)$. However, one can easily check 
that $\varphi(\q,0)\!=0\!$ 
implies that the time averaged density~$\rho(\x)$ given 
by~(\ref{avrho}) is invariant under~(\ref{red}). As $\rho(\x)$ is 
singular at the origin,  so has to be the effective 
density $\rho_{eff}¥(\x,t)$,  although the singularity 
may be more   spread in time if   the assumption of 
time-reversal symmetry is relaxed.

\section{Discussion}

In this paper we have confirmed the old suggestion~\cite{sal,mitp,mit}
 that highly excited 
strings behave like random walks. We have done this by extracting   
---from the appropriate averaged inclusive  cross 
sections---  
probabilities for joint detection of the target string  at two different space-time 
points. These probabilities can be written as a 
convolution~(\ref{proconv})  of closed random walks and  provide, not just a static 
picture of the ensemble of excited closed strings, but also a description of 
its time evolution.

The form of the
joint probability~(\ref{proconv}) can also be inferred from 
the semiclassical 
argument presented at the end of section~3. Nevertheless, it is 
satisfying to see  it emerge from the full-fledged string computation 
in~\cite{ff} after a set of well controlled 
approximations\footnote{The excellent agreement between the quantum 
string computation and the semiclassical argument indicates that 
typical highly excited strings admit a semiclassical description 
 almost down to the string scale. Non-typical states may, or may not,  
 admit such a description.}. In fact, one could  turn the argument around and 
use semiclassical information to obtain effective form factors and 
inclusive cross sections for highly excited strings, 
without ever doing  a detailed string computation.
Indeed, inverting~(\ref{cor})  and writing the partially integrated 
correlator in terms of probabilities yields
\beq\label{use}
|\F_{NN'}(q^2)|^2={1\over T}\int_0^{T}dt\int d^dx\, e^{-i(\q\,\cdot \x-\om t)}\,
P_{st}(\x\,, t)\;.
\eeq  

This formula is useful because the probability in the r.h.s. can be 
written directly for any state  admitting a semiclassical 
interpretation. Such is the case, for instance, for the long lived 
string states considered in~\cite{ang,semi,long,type2,search}. The 
point is that~(\ref{cor}), and therefore~(\ref{use}), are valid not 
just for the mixed state represented by the ensemble 
average~(\ref{ens}), but for any initial target state, mixed or pure.
Indeed, one can easily check that the argument leading to~(\ref{cor}) goes 
through as long as the ensemble average~(\ref{ens}) is replaced
by
\beq
\la A\ra\equiv \mathrm{Tr} (\hat\rho A)
\eeq
where $\hat\rho$ is the appropriate density matrix ---not to be 
confused with the spatial densities considered in this paper!
In other words, as long as one considers  inclusive processes 
where the \emph{final} state of the target is not measured, one ends up with 
a formula relating the cross section to an averaged density-density 
correlator. Considering different initial states only affects the type 
of average.

Nevertheless, one should be aware that~(\ref{use}) is valid only for 
(relatively) small momentum transfer. This is not surprising, since 
any geometric description of  string states is expected to break down at the 
string scale. Less obvious is the requirement of high energy probes, 
which is necessary for the validity of the factorization 
property~(\ref{closedregge}) of the averaged rate~\cite{ff}. The reason may 
be that, in the Regge limit of  high 
energy   and fixed $q^2$,  the scattering is dominated by $t$-channel exchange, 
whereas at lower energies $s$-channel capture and re-emission as 
thermal radiation probably plays an important role.

One possible generalization of the results presented here is the 
extension to superstring inclusive cross sections. There one should face the problem 
of defining what is meant by a `super random walk'. As far as we know, 
no detailed  generalization of the random walk model has been worked 
out for the superstring. One obvious problem is the fact that the 
occupation numbers for fermionic oscillators can never be large, and 
it is not clear what  a semiclassical description means in that 
case.
One clue may be provided by the computation of the massless emission 
rate for a very long lived superstring state in~\cite{search}. The 
correct rate for the dominant decay channel is obtained by a 
classical computation that includes only  bosonic degrees of freedom.
It would also  be interesting to use the random walk picture developed in 
this paper to compute the decay properties of typical highly excited 
strings. The results could be compared with those obtained by 
oscillator methods in~\cite{ayr,dec} for the bosonic string, and 
in~\cite{search,bin} for the superstring.

Throughout this paper we have considered two different 
densities. On the one hand we have the string density $\rho_{st}$ that represents 
the actual string distribution and is regular and time-independent 
---even though its
 two point correlators are  time-dependent. On the other 
hand is the effective density $\rho_{eff}$, that turns out to be both 
singular and time-dependent. 
The effective density $\rho_{eff}$  acts as the source 
for  the classical potential that reproduces the inclusive cross 
sections in the Born approximation, but does not  describe 
the actual string distribution. By the same token, the string density 
$\rho_{st}$ can not be considered as the source for the classical 
background.

In this regard, we would like  to stress the role played by decoherence.
Indeed, both the random walk picture and   the form of the effective 
density~(\ref{fourho}) are valid only for the ensemble considered 
here, i.e., for the density matrix
\beq
\hat\rho={1\over\G_c}\sum_{\Phi_{i}|N}|\Phi_{i}\ra\la\Phi_{i}|\;.
\eeq
Other initial states, mixed or pure, will give rise to different 
classical backgrounds. This is reminiscent of the situation in the 
D-brane approach to black holes. As stressed in~\cite{mye,ama}, in 
order to get the correct black hole properties one has to consider a 
decoherent ensemble of all degenerate D-brane configurations. Less 
generic configurations in the D-brane side give rise to geometries 
that cannot be interpreted as black holes~\cite{math1,math2,math3}.
On the other hand, it is the decoherence associated 
with not measuring the \emph{final} state of the target in an  inclusive cross 
section that 
seems  to be responsible for the striking differences between $\rho_{st}$ and 
$\rho_{eff}$ in our case.

In the D-brane approach one usually computes  physical properties 
such as entropies or grey-body factors at weak coupling, and 
compares the results with those obtained by considering a strongly 
coupled  general relativity solution. Even though  singularities and horizons 
appear only on the gravity side, one usually assumes 
that, as the coupling is increased, the microscopic string 
configuration  undergoes a collapse and becomes the source for the 
black hole geometry~\cite{khuri,cor,hor,dam}. The example 
studied in this paper suggests a different possibility, more in line 
with~\cite{comp}: In our computations, both the string system 
and the classical 
background are at weak coupling, and yet the corresponding 
densities $\rho_{st}$ and $\rho_{eff}$ are very different. As 
mentioned above, one is static 
and
regular, and can not even be considered as the source for the 
background potential; the other is singular and describes a sequence of collapses 
and `big-bangs'. So, maybe the strings never collapse, 
and collapse and `no-collapse' are just 
complementary views of the same reality.

\begin{acknowledgments}
It is a pleasure to thank  I.L.~Egusquiza, R.~Emparan, K.~Kunze, 
M.~Tierz, M.~Uriarte, 
  M.A.~Valle-Basagoiti and  M.A.~V\'azquez-Mozo  for  useful and
interesting discussions. I feel  specially indebted to Valle-Basagoiti for 
discussions that were of great help in clarifying the relation 
between cross sections and structure functions presented in section~3. 
This work  has been  supported in part by
the Spanish Science Ministry under Grant FPA2002-02037 and by  University of 
the Basque Country Grant UPV00172.310-14497/2002. 
\end{acknowledgments}

\appendix

\section{Computation of the string density }

In this appendix we formulate the (fixed time) sum over closed random walks as a 
euclidean path integral, and use it to compute the string density 
$\rho_{st}$. The probability that a string with free center of mass 
goes \emph{simultaneously} through two points separated by an interval $\Delta \x$ is given 
by~(\ref{free}). Writing 
\beq\label{two}
\exp\bigg(-{M\over 
2T_H}{\Delta\x\,^2\over \s(4\pi M-\s)}\bigg)=\exp\bigg(-{1\over 
8\pi T_H}{\Delta\x\,^2\over \s}\bigg)\exp\bigg(-{1\over 
8\pi T_H}{\Delta\x\,^2\over \s(4\pi M-\s)}\bigg)
\eeq
suggests the following form for the path integral
\beq\label{path}
{\cal I}_{0}¥=\int{\cal D} \X\,e^{-S_0[\X]}\;\;\;\;,\;\;\;\; S_0[\X]={1\over 8\pi 
T_H}\int_0^{4\pi M} d\s \Bigg({d\X\over d\s}\Bigg)^2
\eeq
subject to the periodicity condition $\X(4\pi M)=\X(0)$. 
Indeed, this is just the generalization of~(\ref{two}) to a path 
that has to go through an infinite number of points in the continuum 
limit. 

As a check of~(\ref{path}), consider the sum over 
all periodic paths that go through two points separated by $\Delta \x$.
As $S_0[\X]$ is quadratic, it is enough to compute the 
classical solution to $\X''=0$, subject to the constraints 
$\X(0)=\X(4\pi M)=0$ and $\X(\s)=\Delta \x$. This is given by
\beq
\X_{cl}¥\,(\s')=\left\{ \begin{array}{ll} \big({\s'\over\s}	
\big)\Delta\x\ & \;\;\; 0\leq \s'\leq \s\\ & \\ \big({\s'-4\pi M\over \s-4\pi 
M}\big)\Delta\x & \;\;\;\s<\s'\leq 4\pi M \end{array} \right.
\eeq
The action for this classical path is 
\beq
S_0[\X_{cl}]={1\over 8\pi T_H}\int_0^{4\pi M} d\s' \X_{cl}^{\,'2}={M\over 
2T_H}{\Delta\x\,^2\over \s(4\pi M-\s)}
\eeq
which agrees with the exponent in~(\ref{two}).

In order to compute $\rho_{st}$ for an ensemble of strings with center 
of mass at the origin, the action has to be modified so that the path integral
 incorporates this condition. This is achieved by defining
\beq
{\cal I}_{1}¥=\int{\cal D} \X\,e^{-S_1[\X]}\;\;\;\;,\;\;\;\; S_1[\X]=S_0[\X]+{1\over 
4\pi T_H} \int_0^{4\pi M}d\s\, \vec\mu\cdot \X\,(\s)
\eeq
where $\vec 
\mu$ is a $\s$-independent Lagrange multiplier that enforces the constraint $\X_{cm}=0$.
Now, consider the sum over all periodic paths ---with center of mass at 
the origin--- that go through a point $\x$. We can use invariance 
under translations in $\s$ to set $\X(0)=\x$.
As before, the path integral is 
determined by the classical solution to the equation of motion 
$\X\,''=\vec \mu$,
\beq
\X_{cl}¥\,(\s)=\vec a+\vec b \,\s+\med\,\vec \mu\, \s^2\;.
\eeq
Imposing $\X_{cl}¥(0)=\X_{cl}¥(4\pi M)=\x$ yields $\vec a=\x$, $\vec b=-2\pi 
M\vec \mu$. Then,
\beq
\int_0^{4\pi M}d\s\,  
\X_{cl}¥\,(\s)=0\;\;\Longrightarrow\;\;\vec\mu={3\x\over 4\pi^2M^2}
\eeq
and
\beq
S_{1}¥[\X_{cl}]={1\over 8\pi T_H}\int_0^{4\pi M} d\s \X_{cl}^{\,'2}=
{3\x\,^2\over8\pi^2MT_H}\;.
\eeq
This, together with the normalization condition, gives~(\ref{rhost}) 
uniquely.


\begin{thebibliography}{99}


\bibitem{sal} P.~Salomonson and B.~Skagerstam, \textit{On Superdense 
Superstring Gases: A Heretic String Model Approach}, Nucl. Phys. {\bf B268} 
(1986) 349.

\bibitem{mitp} D.~Mitchell and N.~Turok, \textit{Statistical 
Mechanics of Cosmic Strings}, \prl{58}{1987}{1577}. 

\bibitem{mit} D.~Mitchell and N.~Turok, \textit{Statistical 
Properties of Cosmic Strings}, Nucl. Phys. {\bf B294} (1987) 
1138.

\bibitem{thor1} D.A.~Lowe and L.~Thorlacius, Phys. Rev, {\bf D51} (1995) 
665 [hep-th/9408134].

\bibitem{thor2} S.~Lee and L.~Thorlacius, Phys. Lett. {\bf B413} (1997) 
303 [hep-th/9707167].

\bibitem{khuri} R.S. Khuri, \textit{Self-Gravitating Strings and 
String/Black Hole Correspondence}, \plb{470}{1999}{73} [hep-th/9910122].

\bibitem{cor} G.T.~Horowitz and J.~Polchinsky, \textit{A 
Correspondence Principle for Black Holes and Strings}, Phys. Rev. {\bf D55} 
(1997) 6189 [hep-th/9612146].

\bibitem{hor} G.~T.~Horowitz  and J.~Polchinski, \textit{Self 
Gravitating Fundamental Strings}, \prd{57}{1998}{2557}  
[hep-th/9707170].

\bibitem{dam} T.~Damour and G.~Veneziano, \textit{Self-gravitating 
Fundamental Strings and Black Holes}, Nucl. Phys. {\bf B568} 
(2000) 93 [hep-th/9907030].

\bibitem{pepe} J.L.F. Barb\'on and E. Rabinovici, \textit{Touring the 
Hagedorn Ridge}, hep-th/0407236.

\bibitem{thorn} O.~Bergman and C.~B.~Thorn, \textit{String Bit Models 
for Superstring}, \prd{52}{1995}{5980} [hep-th/9506125].

\bibitem{brane} E.~Halyo, A.~Rajaraman and L.~Susskind, 
\textit{Braneless Black Holes}, \plb{392}{1997}{319} [hep-th/9605112].

\bibitem{halyo} E.~Halyo, \textit{Gravitational Entropy and String 
Bits on Stretched Horizons}, hep-th/0308166.

\bibitem{rama} S. Kalyana Rama, \textit{Size of Black Holes through 
Polymer Scaling}\plb{424}{1998}{39} [hep-th/9710035].

\bibitem{had} L.~Susskind, \textit{Structure of Hadrons implied by 
Duality}, \prd{1}{1970}{1182}.

\bibitem{size} M.~Karliner, I.~Klebanov and L.~Susskind, \textit{Size 
and Shape of Strings}, Int. J. Mod. 
Phys. {\bf A3} (1988) 1981.

\bibitem{lor} L.~Susskind, \textit{Strings, Black Holes and Lorentz 
Contraction}, \prd{49}{1994}{6606} [hep-th/9308139].

\bibitem{comp} L.~Susskind, \textit{String Theory and the Principle 
of Black Hole Complementarity}, \prl{71}{1993}{2367} [hep-th/9307168].

\bibitem{bo} D.~Mitchell and B.~Sundborg, \textit{Measuring the Size 
and Shape of Strings}, Nucl. Phys. {\bf B349} (1991) 159. 


\bibitem{ff} J.L. Ma\~nes, \textit{String Form Factors}, 
\jhep{01}{2004}{033} [hep-th/0312035].

\bibitem{gsw} M.~Green, J.~Schwarz and E.~Witten, \emph{Superstring Theory}, 
Vols. I and II, Cambridge~1987.

\bibitem{pol} J.~Polchinski, \emph{String Theory}, Vols. I and II, 
Cambridge~1998.

\bibitem{par} G.~Parisi, \textit{Statistical Field Theory}, New York~1988.

\bibitem{grav} C.W. Misner, K.S.~Thorne and J.A.~Wheeler, 
\textit{Gravitation}, W.H. Freeman and Co., San Francisco 1973.

\bibitem{ang} R.~Iengo and J.~G.~Russo, \textit{The Decay of Massive 
Closed Superstrings with Maximum Angular Momentum}, 
\jhep{11}{2002}{045} [hep-th/0210245].

\bibitem{semi} R.~Iengo and J.~G.~Russo, \textit{Semiclassical Decay 
of Strings with Maximum Angular Momentum}, \jhep{03}{2003}{030} 
[hep-th/0301109].

\bibitem{long}  D. Chialva, R.~Iengo and J.~G.~Russo, \textit{Decay of Long-Lived 
Closed Superstring States: Exact results} \jhep{12}{2003}{014} [hep-th/0310283].

\bibitem{type2}  D. Chialva and R.~Iengo, \textit{Long-Lived 
Large Type II Strings: decay within compactification} 
\jhep{07}{2004}{054}  [hep-th/0406271].

\bibitem{search}  D. Chialva, R.~Iengo and J.~G.~Russo, \textit{Search 
for the most stable massive state in superstring theory},  
hep-th/0410152.
  
\bibitem{ayr} D.~Amati and J.G.~Russo, \textit{Fundamental Strings as 
Black Bodies},  Phys. Lett. {\bf B454} (1999) 
207 [hep-th/9901092].

\bibitem{dec} J.~L.~Ma\~nes, \textit{Emission Spectrum of Fundamental 
Strings: An  Algebraic Approach}, \npb{621}{2002}{37} [hep-th/0109196].

\bibitem{bin} B. Chen, M. Li and J.-H. She,  \textit{The fate of massive F-strings}, hep-th/0504040.

\bibitem{mye} R.~Myers, \textit{Pure States don't wear Black}, Gen. Rel. Grav. {\bf 29} (1997) 1217, 
[gr-qc/9705065].

\bibitem{ama} D.~Amati, \textit{Black Holes, String Theory and 
Quantum Coherence}, hep-th/9706157.

\bibitem{math1} S.D. Mathur, A. Saxena and Y.K. Srivastava, 
\textit{Constructing hair for the three charge black hole}, \npb{689}{2004}{415} 
[hep-th/0311092].

\bibitem{math2} S. Giusto, S.D. Mathur and  A. Saxena, 
\textit{Dual geometries for a set of 3-charge microstates}, 
hep-th/0405017.

\bibitem{math3} S.D. Mathur, \textit{Where are the states of a black 
hole?}, hep-th/0401115.


\end{thebibliography}
\end{document}